# Physical Origin of H-Mode

Kwan Chul Lee

Korea Institute of Fusion Energy, Daejeon 34133, Korea

*Correspondence to Kwan Chul Lee : (e-mail: kclee@kfe.re.kr)

**Abstract**

**The high confinement mode (H-mode), the most important operation mode for the nuclear fusion reactor, has been studied for 42 years, but the transition mechanism has not been unanimously agreed so far. Four decades of H-mode experiments revealed many features of heating power threshold ($P_{th}$) for the low to high confinement (L-H) transition, where $P_{th}$ is proportional to the toroidal magnetic field (B), inversely proportional to the ion mass ($m_i$), and $P_{th}$ has U-shaped dependence on the plasma density. It is found for the first time that this U-shaped rollover dependence came from $P_{th}$ is inversely proportional to the multiplication of plasma density ($n_i$) and the squire of neutral density ($n_n$). The reason for the neutral density involved in the L-H transition is that the turbulence suppression takes place by the viscous force of the ion-neutral friction. When the plasma is in the equilibrium by the compensation of turbulence-induced return current to the gyro-center shift current generated by the ion-neutral charge exchanges, the Reynolds number ($Re$), the ratio of inertial force to the viscous force, can predict the onset of laminar to turbulent flow. $Re$ for the plasma-neutral interaction is proportional to $B_T / (m_i\, n_i n_n^2)$ which is well agreed with the experimental results of $P_{th}$. $Re$ is also proportional to the second gradient of the radial electric field which is agreed with experiments. 15 characteristics of L-H transition are explained by $Re$ including the feature of $P_{th}$ having favorable dependence on the ion grad B drift toward the x-point.**

I.    Introduction

Among important unsolved physics problems, it can be found that the most important challenge in plasma physics is "What is the physical origin of the H-mode?", which is regarded as on the same level as "Is it possible to produce energy from nuclear fusion?". The idea that unlocking the secret of H-mode could lead to commercial nuclear fusion is not far-fetched. Nuclear fusion energy utilizes the phenomenon that energy is generated by the mass defect that occurs when the nuclei of, for example, deuterium and tritium fuse and convert into helium nuclei and neutrons according to the equation $E=MC^2$. Since atomic nuclei are positively charged, they need to be heated to more than 100 million degrees to make them stick together, and when they are that hot, they become plasma, separated into positively charged ions and negatively charged electrons instead of neutral atoms, and there is no container to hold such hot plasma, so it uses the helical motion of charged particles in a magnetic field. Plasma particles spiral and become trapped in the magnetic field, and if the magnetic field is shaped like a doughnut, they can only spin and move within it, so become trapped. Such a torus-shaped plasma confinement device is a tokamak.

The scientists who developed the atomic bomb in 1945 in the Manhattan Project were devastated by its destructive power, and Fermi, the best of them, completed the first fission reactor only six years later. By this time, Teller had developed a hydrogen bomb that used both fission and fusion energy, leading people to believe that a fusion reactor would be ready within a decade. Project Matterhorn was a plan for the peaceful use of fusion energy and was developed at Princeton University's Plasma Physics Laboratory. But that's where physics encounters an unexpected obstacle. The magnetic field cannot completely trap the plasma, as particles trapped in the inner magnetic field and particles trapped in the outer magnetic field can change their positions when they collide, and a small amount of plasma will escape the magnetic field from a higher density to a lower density, which, according to classical physics, is inversely proportional to the square of the magnetic field, so it should rapidly decrease when a strong magnetic field is applied. However, in the experiments, it is found that the plasma leaking is only inversely proportional to the first power of the magnetic field, and the amount of escape is more than 100 times larger than the expected value. This phenomenon was proportional to Bohm diffusion equation, which was also discovered by experiment, so its physical origin was uncertain. This unexplained diffusion phenomenon called 'anomalous transport' had to be studied, while 30 years passed far from the success of nuclear fusion. Then came the good news from the German fusion device ASDEX. H-mode had been discovered.

## II. What is H-mode?

H-mode refers to the high confinement mode (H-mode) relative to the low confinement mode (L-mode) when the power of the heating device that raises the temperature of the plasma is increased by a certain amount, resulting in a transition to a higher confinement state than the existing low confinement. In 1982, it was discovered for the first time that even a small increase in heating power reduces the amount of plasma that escapes, increasing the temperature and density, which doubles the plasma pressure, and normally the higher the temperature, the shorter the time of confining particles, but in H-mode, the confinement time is about twice as long as in L-mode despite the increase in temperature [1]. Since then, H-mode has occurred in almost all tokamaks around the world, and various characteristics of H-mode have been observed experimentally. The main characteristics are summarized below.

(1) H-mode only occurs when the heating power is increased beyond a certain point.

(2) In the L-mode, turbulence exists at the edge of the tokamak, but when H-mode occurs, the turbulence at the edge is suppressed and a transport barrier is formed at that location.

(3) The threshold power ($P_{th}$) required for the transition from L-mode to H-mode (L-H transition) increases with the magnetic field.

(4) $P_{th}$ decreases with the mass of the ions comprising the plasma. For example, a deuterium plasma has half the $P_{th}$ of a hydrogen plasma.

(5) H-modes do not occur well if there are too many impurities. (Therefore, wall cleaning to reduce impurities is important for H-mode generation.)

(6) $P_{th}$ increases when the density of the plasma is too low or too high (there is a U-shaped dependence).

(7) H-mode occurs when the right amount of neutrals are inserted.

(8) H-modes are more likely to occur in diverter plasmas, where the plasma edge hits the wall near the x-point of the magnetic field configuration at the bottom or top of the tokamak, and less likely in limiter plasmas, where the plasma hits the inner wall around midplan.

(9) $P_{th}$ is lower when the direction of drift velocity of plasma ions due to the magnetic field gradient is toward x-point than away from x-point.

(10) L-H transition has a bifurcation and a hysteresis phenomenon, that is, once the transition occurs, it goes backward transition differently from the process of the forward transition.

(11) The radial electric field ($E_r$) near the boundary increases after the L-H transition.

(12) The profile of $E_r$ plays an important role in L-H transition (H-mode can also be generated by artificially applying an electric field).

(13) The L-H transition takes place in a short time, about 30~100 microseconds.

(14) H-mode seems random in its approach, sometimes occurring and sometimes not, even under superficially identical conditions.

(15) The H-mode is most often followed by an edge localized mode (ELM), which recurs periodically.

In the 42 years since the discovery of the H-mode, numerous researchers have published numerous papers trying to answer the question of the physical origin of the H-mode, but as mentioned in the introduction, it remains a major unsolved problem. The answer to this question should be able to explain all the 15 characteristics listed above. The solution in this paper is based on 42 years of experimental research, but it is largely unrelated to existing other theoretical works of plasma physics, so we will not introduce existing theories, but rather recall Feynman's famous quote: "It doesn't matter how beautiful your theory is, it doesn't matter how smart you are. If it doesn't agree with experiment, it's wrong."

III.     Current and electric field due to gyro-center shift

Collisions between plasma particles were already mentioned for the phenomenon of plasma particles escaping from a magnetic field without being completely trapped in it. Inside the plasma, there is a small amount of neutral particles (mainly atoms) that have not yet been ionized, and their density increases towards the boundary of the plasma. When ions in a plasma trapped in a magnetic field collide with neutrals, their positions change. This phenomenon is mainly caused by the charge exchange reaction, which is a typical ion-neutral reaction, and as shown in figure 1 below, when an ion that has been in gyro-motion due to the magnetic field meets a passing neutral, the electron bound in the neutral is transferred to the ion, and the particle that was previously an ion becomes a new neutral, and the neutral becomes an ion.

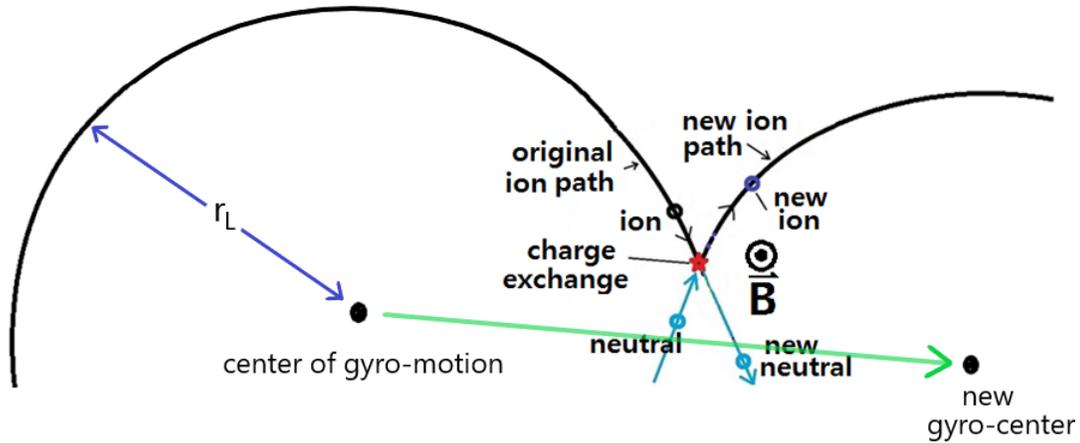

**Figure 1 Shifting the center of a gyro-motion by a charge exchange reaction. The magnetic field is in the direction from the ground.**

As a result of this reaction, the gyro-centers of the ions are shifted. This phenomenon also occurs in the collision of an electron with a neutral, but since the mass of the electron is thousands of times smaller than that of the ion, the gyro-radius is also smaller and electron gyro-center shift becomes negligible compared to the ion gyro-center shift. If the relative velocity between ions and neutrals is isotropic, then the gyro-center shifts will also be isotropic, canceling each other out, but there are two cases where they are not isotropic. The first is when the neutrals are moving in one direction, such as in the equatorial electro jet (EEJ), but at the boundary of the tokamak, the direction of neutral motion is largely isotropic, especially in the perpendicular component to the magnetic field, so this component is not relevant. The second is when the ion is moving in one direction due to drift velocity. The first drift velocity is due to the pressure gradient of the ions; if the plasma pressure on the left side of figure 1 is greater than the right side, the average velocity of the ions coming down from the top is greater than the average velocity of the ions coming up from the bottom, so the gyro-center shift to the right side occurs as shown in the figure. The second drift velocity is due to the density gradient of neutrals; if the density of neutrals is low on the left side of the figure and high on the right side, the ions have a higher probability of colliding with neutrals from the top to bottom than from the bottom to top, so the gyro-center shift to the right will occur. If the magnetic field on the left side of the figure is higher than the right side, the gyro-radius will be larger on the right side and smaller on the left during the ion's gyro-motion, resulting in an overall downward drift; this is the grad-B drift (B is the magnetic field). There can be other drift velocities, but the fourth and most important one is electric field driven drift. If the electric field is directed from right to left in the figure, the ion will be decelerated during its gyro-motion to the right and accelerated to the left, resulting in an overall upward drift. This is called the ExB drift velocity (E is the electric field ). Therefore, at the boundary of the plasma where the plasma and neutral gas meet, there is a region where the two substances overlap and many charge exchange reactions occur, and in this region, the pressure of the plasma and the density of the neutrals change rapidly, so there is a one-directional drift velocity due to the plasma pressure gradient and the neutral density gradient, and as a result, the ions always generate a current by a gyro-center movement from the plasma region to the neutral region, which is called the gyro-center shift (GCS) current; $J^{GCS}$ [Appendix 1]. The GCS current is expressed by the following equation [2].

$$J^{GCS} = \frac{n_i m_i v_{i-n} \upsilon_{i-n}}{B} \quad (1)$$

where $n_i$ is ion density, $m_i$ is ion mass, $v_{i-n}$ is ion-neutral collision frequency, and $\upsilon_{i-n}$ is the relative velocity between the ion and neutral, which is the sum of all drift velocities.

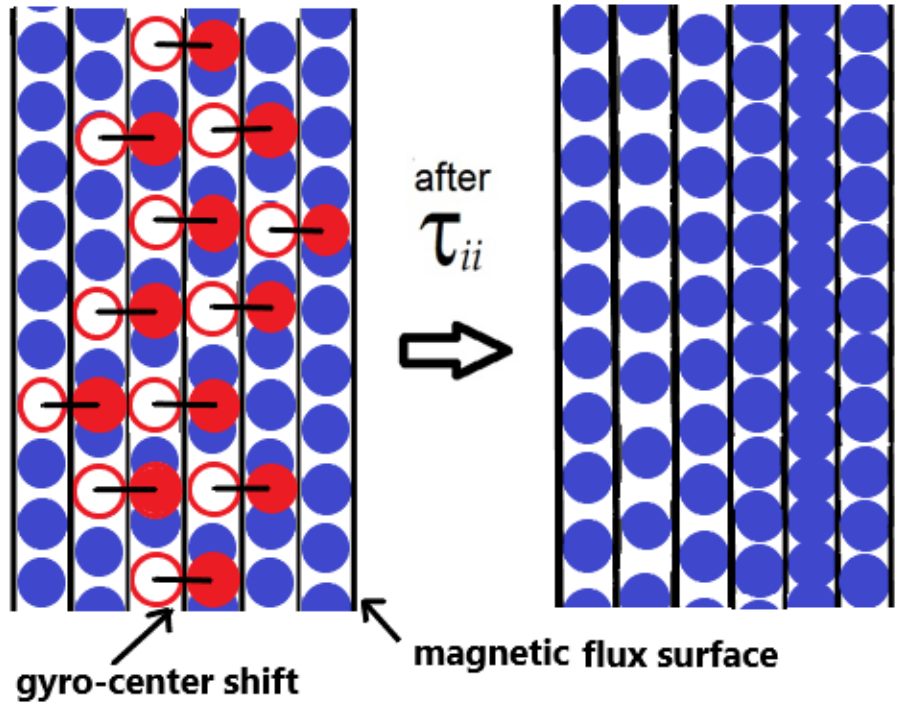

**Figure 2 Separation of charges by GCS current; As shown in the left side of the figure, the ions trapped in the magnetic field from the bottom to the top make a gyro-center shift in one direction, and after the equilibrium time (ion-ion collision time) when the particles in the same magnetic flux are equalized, the distribution of ions becomes higher on the right side, as shown in the right side of the figure. This process is according to the continuity equation; $\nabla \cdot J = -\frac{\partial \rho}{\partial t}$, (here $\rho$ is the charge density). And an electric field is generated from positive on the right to negative on the left by $\nabla E = \frac{\rho}{2\epsilon_0}$, ( $\epsilon_0$ is permeability of air).**

$J^{GCS}$ causes the separation of charges over time by the continuity equation, as shown in figure 2, as the magnitude of $J^{GCS}$ varies with radius at the boundary of the tokamak [3]. The separation of charges by $J^{GCS}$ always has the direction of pushing positive ions toward the edge from the inside of the plasma regardless of the direction of the magnetic field, and when an electric field is formed, the resulting ExB drift velocity cancels the other drift velocity caused by the plasma pressure gradient and neutral density gradient, and the charge separation stops and reaches equilibrium. The electric field generated by this GCS current forms a radial electric field at the boundary of the tokamak that is always directed toward the inner of the tokamak and is the source of the electric field found in low-temperature plasmas such as the cathode spot of an arc discharge or the black aurora in Earth's ionosphere. An important point to note here is that the ExB drift velocity is not limited to ions, but also applies to electrons, and for a fully ionized plasma such as a tokamak, it makes convection of the plasma because it has the same direction and magnitude for ions and electrons.

IV.   Diffusion by turbulence

Bohm diffusion equation, discovered by Bohm in 1949, describes phenomena in low-temperature plasmas such as arc discharges, which are different from high-temperature plasmas such as tokamaks, but the experimental results of plasma diffusion measured in various devices, including tokamaks, were proportional to Bohm diffusion. That is, the diffusion of plasma trapped in a magnetic field is proportional to the temperature of the plasma and inversely proportional to the magnetic field. However, the front coefficient of 1/16 of Bohm diffusion was not always matched, and it was discovered that this factor becomes smaller as the turbulence magnitude decreases. In 1986, the French tokamak device published the observation that the confinement time, which is inversely proportional to the diffusion, is inversely proportional to the square of the density fluctuation ($\delta n/n$) [4]. Our goal is to determine under what conditions the plasma in a tokamak becomes turbulence, but to do so, we first need to find the diffusion coefficient that results from the presence of turbulence in the plasma. Density fluctuations in tokamaks have been found to be larger at the boundary than in the core of the plasma [5]. In other words, the turbulence that causes diffusion in the plasma is stronger at the boundary where it overlaps with neutrals, so if you look closely at the vortices generated by the turbulence, you will see a structure in which the plasma-rich and neutral-rich regions are mixed by vortices, as shown in figure 3 [6]. The plasma particles are trapped by the magnetic field, but the neutrals are not affected by the magnetic field, so the turbulence moves them around, creating a denser mass of neutrals than the surrounding plasma.

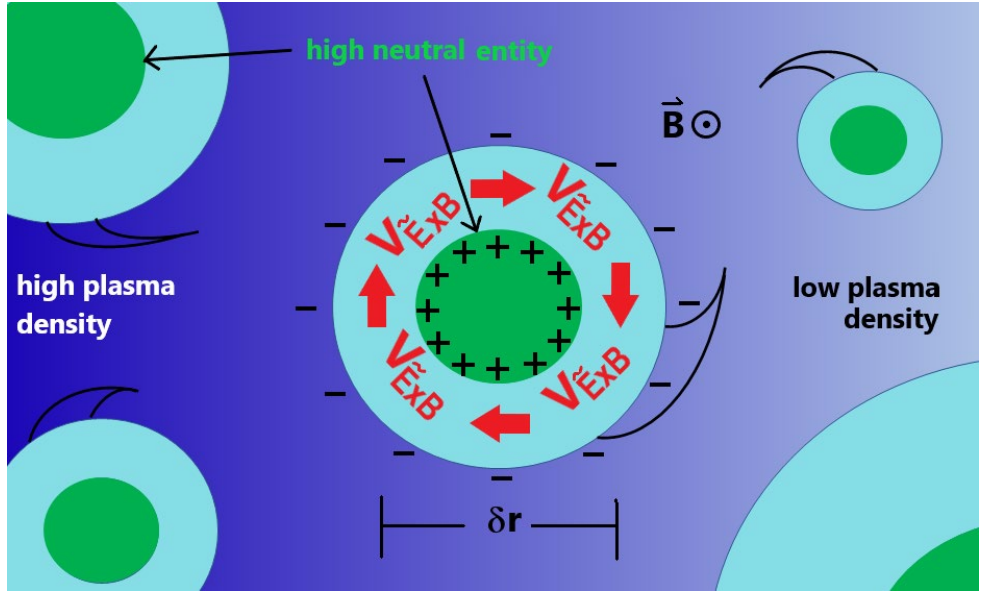

**Figure 3 The formation of micro-vortices caused by turbulence. The left side of the figure has a higher plasma density than the right side.**

The GCS current always expels ions from the plasma to the neutral side, so that positive charges are collected on the neutrals and negative charges are collected on the plasma, as shown in the figure (this phenomenon is microscopic, as it is on a much smaller scale compared to the radial electric field that is generated in the boundary region of the tokamak as a whole). The resulting ExB drift velocity then creates a circularly shaped convection, as shown in the figure, in which both ions and electrons are circularly convected across the magnetic field. Since the plasma density on the left side of the figure is higher than on the right side, this circular convection causes diffusion of the plasma, and the amount of diffusion in one revolution can be calculated as the derivative of the plasma density ($\frac{dn}{dr}$) multiplied by the distance traveled in one revolution (δr), which must be multiplied by the ratio ($\frac{\delta n}{n}$) since not all plasma particles undergo this circular convection, but only the plasma component that participates in this perturbation [Appendix 2]. Multiplying by the velocity of ExB circular convection ($\frac{\tilde{E}}{\pi B}$) to get the amount of diffusion per unit time, the flow rate of the diffusing plasma is $\frac{\delta n}{n}\frac{\delta r}{\pi}\frac{\tilde{E}}{B}\frac{dn}{dr}$. The components of this other than $\frac{dn}{dr}$ are the diffusion coefficient. Here, the magnitude of $\delta r \tilde{E}$ is difficult to know exactly, so we can apply the Boltzmann relation ($kT_e \delta n_e = n_e e \delta \phi$), which states that electrical and thermal energy are in equilibrium when there is a perturbation in the plasma. Where $kT_e$ is the thermal energy (temperature) of the electron, e is the charge of the electron, and $\delta\phi$ is the electric potential perturbation. Since the electric potential is the product of the electric field and the distance, $\delta\phi \approx \frac{\delta r}{2}\tilde{E}$ in the case shown in figure 3, the final diffusion coefficient due to turbulence is

$$D = \frac{2}{\pi}\left(\frac{\delta n}{n}\right)^2 \frac{kT_e}{eB}. \qquad (2)$$

If $\frac{\delta n}{n}$ is about 0.31, then $D = \frac{1}{16}\frac{kT_e}{eB}$, which is the same as Bohm diffusion, but this is a coincidence. Eq.(2) not only explains why the diffusion in fusion devices is proportional to temperature and inversely proportional to magnetic field, but also explains the experimentally found phenomenon that the diffusion in fusion devices is proportional to the square of the density fluctuation. And Eq.(2) can be applied not only to the diffusion of plasma particles, but also to the diffusion of charge density arising from the imbalance of electrons and ions.

V. Reynolds number distinguishes laminar from turbulent flow

H-mode is induced by the turbulence suppression at the boundary of a tokamak. In non-plasma neutral fluid mechanics, there is Reynolds number that distinguishes between laminar and turbulent flow when there is friction in the flow of a fluid. The Reynolds number is defined as the ratio of the inertial force to the viscous force, and if this ratio is lower than about 2300 (generally between 2000 and 4000 is called the transition region, and the exact critical Reynolds number is still the subject of research), the flow is laminar, and if it is higher, it is turbulent. For example, in the flow of water from a faucet, the Reynolds number is proportional to the velocity of the water, so it is laminar at low speeds and transitions to turbulent above a certain velocity. Since inertial force is a force that makes it rotate and viscous force is a force that sticks, it is natural that turbulence occurs when the inertial force is thousands of times greater than the viscous force. This phenomenon can be compared that if you tackle on a walking person, he will collapse in place, but if you tackle on a running person, he will be rolling over. The transition from turbulent to laminar flow in H-mode generation can be characterized by the Reynolds number for the friction between the plasma and neutrals when the plasma moves as a fluid under the influence of electromagnetic forces while the neutrals are not affected by electromagnetic forces. However, the Reynolds number between plasma and neutrals under a magnetic field differs from that of a typical neutral fluid in at least three ways. First, the object causing the friction is a solid, such as a water pipe in the case of tap water, but in the case of tokamaks, it is the neutrals, so it must be defined from an individual particle perspective. The second, dimension of scale, such as the diameter of a water pipe, is reduced to the gyro-radius of an ion in a tokamak. This is because, for a given neutral, an ion that is further away than the gyro radius in the direction perpendicular to the magnetic field cannot collide with that neutral. The third is that the rate at which the fluid experiences friction (the relative velocity of the ions and neutrals) is determined by the equilibrium conditions that arise in the interaction with turbulence. The Reynolds number of a tokamak plasma is expressed as

$$Re = \frac{n_i m_i v_{i-n}^2 / r_L}{n_i m_i \nu_{i-n} v_{i-n}} \quad . \quad (3)$$

Where $r_L$ is the gyro radius of the ion. Since the ion-neutral collision frequency $\nu_{i-n} = \sigma_{i-n} v_i^{th} n_n$ ($\sigma_{i-n}$ is the collision cross-section which is the probability of a collision occurring, and $v_i^{th}$ is the average velocity due to the thermal motion of the ion), Eq.(3) can be transformed to $Re = \frac{\lambda_{i-n}}{r_L} \frac{v_{i-n}}{v_i^{th}}$. $\lambda_{i-n}$ is the distance ion travels before colliding with a neutral (mean free path; $\lambda_{i-n}=1/\sigma_{i-n} n_n$). This equation explains that as the plasma is heated, $v_i^{th}$ becomes larger and $Re$ becomes smaller, so that below the critical $Re$, turbulence changes to laminar flow and H-mode occurs, and since $r_L = \frac{m_i v_i^{th}}{qB}$, the larger the ion mass, the smaller $Re$ becomes, and the larger the magnetic field, the larger $Re$ becomes. It is also explained that the more neutrals, the smaller $Re$ since $\lambda_{i-n}$ is inversely proportional to the density of neutrals. However, it is difficult to investigate $v_{i-n}$ exactly since it changes depending on the conditions, so it is necessary to use the equilibrium condition that determine $v_{i-n}$. Earlier, it was said that when the J[GCS] separates charges to create an electric field, the resulting ExB drift velocity cancels the other drift velocity that was the driving force of the J[GCS] to reach equilibrium, but this is when turbulence does not exist. In reality, turbulence exists, so when the J[GCS] separates charges, the separated charges return due to diffusion by turbulence. Therefore, the equilibrium condition for J[GCS] to stop separating charges is as follows.

$$J^{GCS} = D \frac{d\rho}{dr} \quad (4)$$

By Eq.(4), it will be no more charge separation and equilibrium is achieved. Therefore, when the turbulence is large, the charge separation in the equilibrium state is only a little, so the ExB drift velocity caused by this cannot cancel the drift velocity caused by the plasma pressure gradient and the neutral density gradient. Substituting Eqs.(1) and (2) into Eq.(4), $v_{i-n}$ can be expressed in terms of other parameters, and the final Reynolds number obtained by substituting them into Eq.(3) is given by

$$Re = \frac{4\epsilon_0}{\pi}\left(\frac{\delta n}{n}\right)^2 \frac{B\nabla^2 E_r}{m_i n_i n_n^2 \sigma_{i-n}^2 v_i^{th}}. \qquad (5)$$

We used that the thermal energies of ions and electrons are similar in the boundary region ($kT_i \approx kT_e$). Based on Eq.(5), we can now describe the development of the H-mode. Since the plasma generated in the tokamak initially has a low temperature and density, Eq.(5) shows that $v_i^{th}$ and $n_i$ are small, so $Re$ is larger than the critical value and remains as turbulence. As the density of the plasma increases and further heating proceeds, $v_i^{th}$ increases, $Re$ near the boundary reaches the critical value, and the turbulence in this region is suppressed. In this way, a laminar flow with suppressed turbulence is formed in a region of about a few centimeters near the boundary of the tokamak, and $\frac{\delta n}{n}$ is significantly reduced in this region. As $\frac{\delta n}{n}$ decreases, the diffusion coefficient in Eq.(2) decreases proportionally to the square of it, the number of plasma particles escaping decreases, and this region becomes a transport barrier. As the diffusion coefficient in Eq.(2) decreases, not only the diffusion of the plasma but also the diffusion of the charge separated by the $J^{GCS}$ decreases, so the equilibrium condition in Eq.(4) changes. The $J^{GCS}$ induced by the pressure gradient of the plasma and the density gradient of the neutrals pushes the positive charge out of the boundary, which was returned by diffusion due to turbulence before the H-mode occurred. But when the turbulence is suppressed by the H-mode, the returning charge is reduced. The $J^{GCS}$ separates more charges and the radial electric field increases. The specific characteristics of the H-mode are explained as follows by reviewing the list shown earlier.

(1) Characterization of the H-mode transition by additional heating: This is explained by $v_i^{th}$ in the denominator of Eq.(5).

(2) The suppression of turbulence in H-mode: This is described by Eq.(5)

(3) The proportionality of $P_{th}$ to the magnetic field: When the magnetic field is stronger, the gyro radius becomes smaller and the number of plasma ions colliding with one neutral is correspondingly smaller, which reduces the proportion of viscous force, then, $Re$ becomes larger and the transition to laminar flow becomes more difficult. Therefore, B is in the numerator of Eq.(5).

(4) $P_{th}$ is inversely proportional to the ion mass: The larger the ion mass, the larger the gyro radius, so $Re$ becomes smaller and the transition to laminar flow becomes easier. $m_i$ is in the denominator of Eq.(5).

(5) Characteristic that H-mode is less likely to occur when there are too many impurities: The charge exchange reaction that occurs with the atom that makes the main ion (in fusion plasma, deuterium or tritium has one electron) is a process in which one electron changes its nucleus, and the gyro-center shift occurs clearly as shown in figure 1, but impurities have multiple electrons, so the addition or loss of one electron does not cause a significant change of the gyro-center. Therefore, the greater the concentration of impurity ions compared to the main ions, the less effective the gyro-centric shift and the weaker the viscous force, making it difficult for H-mode to occur. In other words, the more impurities, the less $n_i$ (the density of main ions) in Eq.(5).

(6) The U-shaped dependence of $P_{th}$ increases when the plasma density is too low or too high: This characteristic is a relatively recent discovery. The first derivation of Eq.(5) was in 2009, when U-shape dependence was not well known. For the 15 years since then, U-shape dependence has been confirmed in almost all tokamaks worldwide. Therefore, the fact that Eq.(5) also explains U-shape dependence can be regarded as an experimental validation that Eq.(5) describes H-mode.

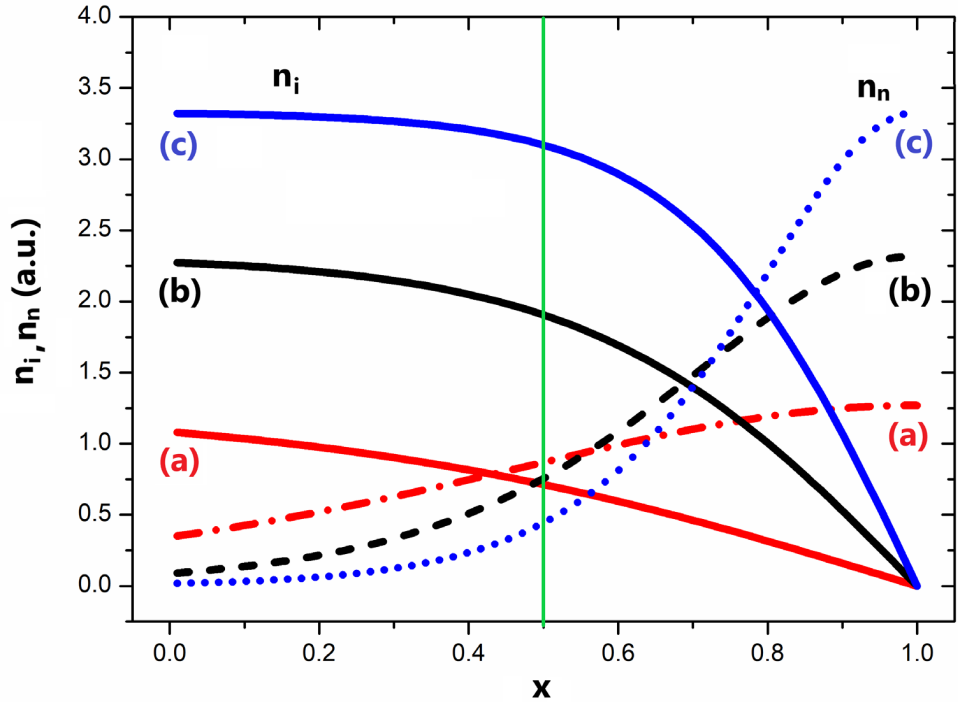

**Figure 4 Profiles of ion density and neutral density as the analytic solution of differential equations ; $n_i(x) = n_{i,core}\tanh(\frac{x}{x_0})$, $n_n(x) = n_{n,edge}\text{sech}^2(\frac{x}{x_0})$, here $n_{i,core} \propto n_{n,edge}$ and $x_0 \propto \frac{1}{n_{i,core}}$. This analytic profiles are well agreed with the experimental results (Fig.17 of reference 8). Profiles of (a),(b), and (c) correspond points (a),(b), and (c) in figure 5.**

According to Eq.(5), the larger the plasma density, the smaller the *Re*, and therefore the smaller the $P_{th}$, which explains the large $P_{th}$ at low density, but does not seem to explain the phenomenon that $P_{th}$ is large again when the density is too high. However, the U-shaped dependence is due to a hidden variable, which changes as the known variable changes. To increase the density of a plasma experimentally, more neutral gas as the fuel for the plasma, must be injected. This leads to an increase in neutrals, and as they ionize, the density of the plasma increases. It would seem that in order to increase the plasma density, the neutral density would also need to increase, but this is only true for the outer side of the plasma. The opposite happens inner side of the plasma. The equations that describe the densities of plasma and neutrals have the form of hyperbolic functions, as shown in figure 4, because the sink of one is the source of the other. The higher the plasma density, the shorter the penetration length for neutrals to spread into the plasma, so a larger neutral density outside the plasma leads to a smaller neutral density inside the plasma (the mathematical solution of this phenomenon has been introduced in a textbook [7] and has been found in the experimental analysis [8]).

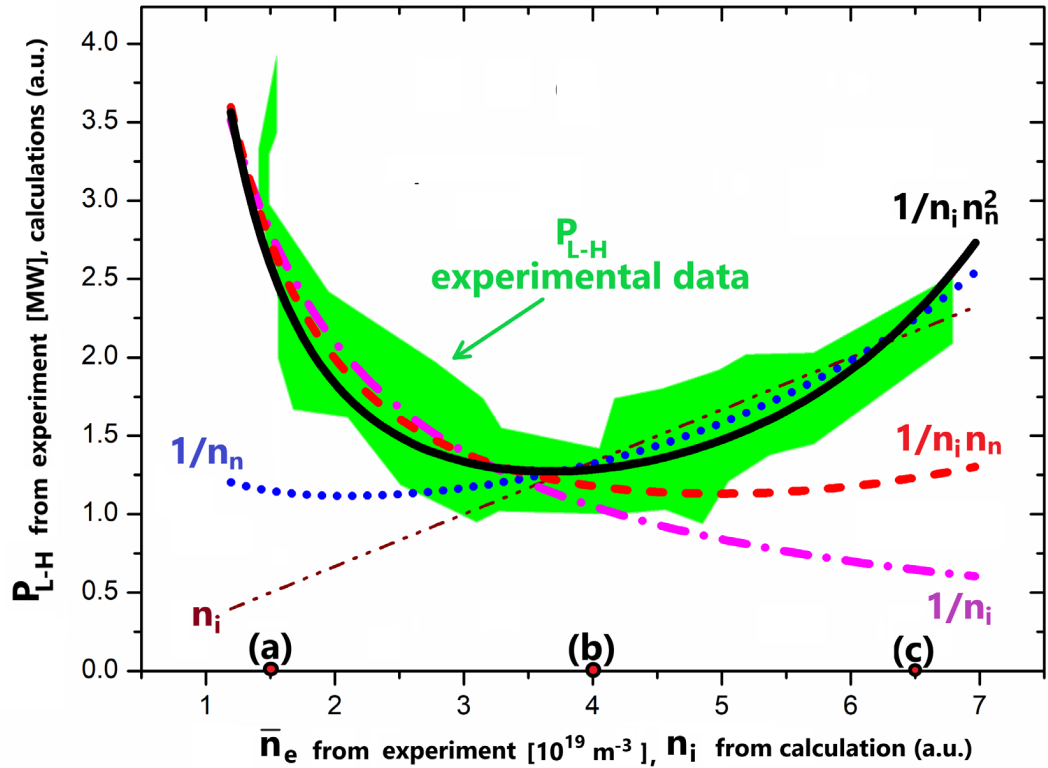

**Figure 5 Comparison of L-H transition power threshold from experiment to the calculations with parameters of $n_i$ and $n_n$. $n_i$ and $n_n$ are taken at $x_0= 0.5$ in figure 4 while the shape is changing from (a) to (c). The arbitrary units came from that the calculations are multiplied by a factor that makes the curve best fit with experimental data. Experimental data are from ASDEX-U [9].**

If we increase the neutral density from (a) => (b) => (c) on the right side of figure 4 to increase the plasma density, the neutral density in the interior of the plasma, where the green line is located, becomes rather small. The dependence of $P_{th}$, including the characteristic that the neutral density decreases as the plasma density increases, is shown in figure 5. The green area represents the experimental measurement, and the combination of the two variables that best fits the experiment is when the $P_{th}$ is inversely proportional to the plasma density multiplied by the square of the neutral density. This is exactly the case of Eq.(5). The turbulence in the plasma is suppressed by the viscous force from the collisions with the neutrals, which only increases when the densities of the plasma and neutrals are simultaneously high, so when the plasma density is very low, the H-mode transition is difficult because there is little plasma, and when the plasma density is too high, the neutral density becomes too low, making the transition difficult.

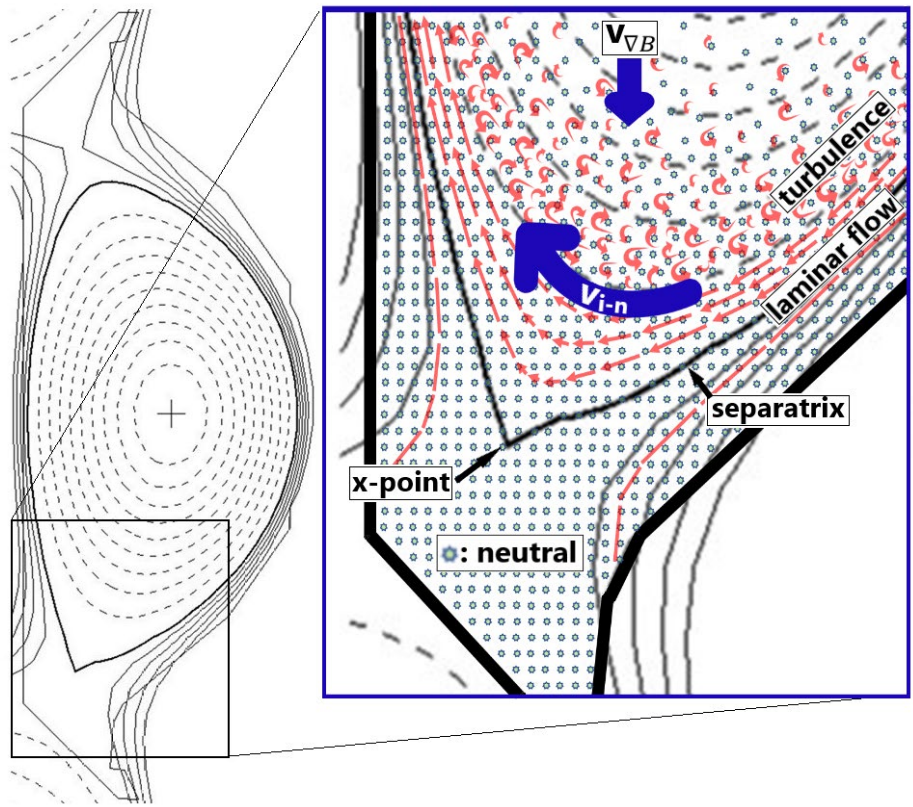

**Figure 6 Schematics of plasma-neutral friction near the x-point. The laminar flow is formed just inside of the separatrix (edge transport barrier) as a result of the L-H transition, in L-mode, the whole region including SOL is covered by turbulence and the overall turbulence magnitude is higher. Even in H-mode, there is turbulence inside of edge transport barrier according to previous calculation [6].**

(7) H-mode occurs when an appropriate amount of neutrals are injected: This phenomenon has been found in H-mode experiments in Japan [10], where the H-mode is more likely to occur when the neutral density is high, which is explained by Eq.(5).

(8) The characteristic that H-mode occurs well in diverter plasma and not in limiter plasma: When the plasma and the wall meet, the plasma particles collide with the wall, the ions and electrons recombine to become neutrals and return to the plasma (recycling). If the surroundings are open like a limiter plasma, the neutrals spread out, resulting in a low density, but in the vicinity of the diverter, as shown in figure 6, the structure of the magnetic field and the wall with the form of an x-point, so the neutrals are easy to be confined, resulting in a higher density. Since the denominator of Eq.(5) has the square of the neutral density, the diverter plasma is more favorable for H-mode generation.

(9) $P_{th}$ is lower when the plasma ion grad B drift velocity direction is toward the x-point than away from the x-point: The ion grad B drift is added to the other ion drift and transferred into the neutral velocity in the charge exchange reaction described in figure 1. Therefore, this velocity component acts to toss the neutrals, so the neutral density becomes higher and the $P_{th}$ becomes lower when the tossing direction is toward the x-point.

(10) In L-H transition, there is a bifurcation and a hysteresis phenomenon: What Eq.(5) represents is the ratio of inertial force to viscous force. However, since the formula for determining turbulence magnitude includes $\frac{\delta n}{n}$ itself, when $Re$ reaches the critical point and turbulence is suppressed, $\frac{\delta n}{n}$ decreases and $Re$ decreases further. This is a hysteresis phenomenon.

(11) The characteristic that the radial electric field ($E_r$) near the boundary increases at the L-H transition: This phenomenon is combined of two processes. The first is the increase in $E_r$ due to the suppression of turbulence, which is caused by the reduction of diffusion-induced return current immediately after the turbulence suppression, as described earlier. The second is the increase in $E_r$ due to the increase in plasma pressure gradient. In L-mode, the diffusion is large, so the plasma pressure is low and the pressure gradient is also small. So the drift velocity of ions proportional to the pressure gradient is small. At this time, the neutral density gradient is also small (figure 4), so L-mode $v_{i-n}$ is smaller than H-mode. Then, since the $J^{GCS}$ in Eq.(1) is smaller, the charge separation is smaller compared to the H-mode. When the transition to H-mode occurs, the diffusion of the plasma decreases and the temperature and density of the plasma increase, which increases the plasma pleasure and the pressure gradient, which increases $v_{i-n}$ and $J^{GCS}$, resulting in more charge separation, which leads to a secondary increase in $E_r$. The increase in $E_r$ due to the turbulence suppression occurs fast, while the increase in $E_r$ due to the pressure gradient increase occurs relatively slowly. This phenomenon is also confirmed in experiments. An increase in $E_r$ and an increase in plasma pressure gradient have been observed together at the L-H transition, but an increase in $E_r$ beyond the increase in pressure gradient has been observed [11].

(12) The profile of $E_r$ plays an important role in L-H transition (H-mode can be generated by artificially applied electric field): Based on this phenomenon, there is a theory that the transition to H-mode is induced by the first derivative of $E_r$ (such as ExB sheering rate). This is because it is observed that the H-mode (transport barrier) occurs at the point in the distribution of $E_r$ where the first derivative term is maximized. However, as shown in Eq.(5), it is the second derivative of $E_r$ that contributes to the H-mode transition. This is because around the inflection point where the second derivative term approaches zero ( $\nabla^2 E_r \approx 0$ ), $Re$ can be smaller than the critical value, even if the other variables in Eq.(5) make it larger. Experimental examples of transport barriers at locations where $\nabla^2 E_r \approx 0$ have been observed in the tokamaks of TFTR [12] and NSTX [13] at the Princeton Plasma Physics Laboratory.

(13) The L-H transition is a fast-occurring phenomenon, on the order of 30~100 microseconds: The transition from L-mode to H-mode involves charge separation and an first step $E_r$ increase by the turbulence suppression. This process is illustrated in figure 2. The change of $J^{GCS}$ by the $E_r$ change can be described by $J^{GCS} = \frac{n_i m_i v_{i-n} E_r}{B^2}$ with replacing $v_{i-n}$ in Eq.(1) as $v_{ExB} \left( = \frac{E_r}{B} \right)$. Combining this with $\nabla \cdot J = -\frac{\partial \rho}{\partial t}$ and $\nabla \cdot E = \frac{\rho}{\epsilon_0}$ to invest the time increase of $J^{GCS}$ as in the form of $J(t) = J_0 e^{\frac{-t}{t_0}}$, leads the time constant of the process $t_0 = \epsilon_0 B^2 / n_i m_i v_{i-n}$ [6], and typical values of $t_0$ at the boundary of a tokamak are tens of microseconds.

(14) Random approach of H-mode: The reader may ask "If Eq.(5) really explains the origin of H-modes, why can't we prove it directly by experiment?" It would be straightforward to measure all the variables in Eq.(5) every time an H-mode occurs and check that $Re$ is a critical value. This has not been possible to date because some of the variables in Eq.(5) are not routinely measurable. The neutral density, for example, has not been measured routinely. It has been calculated indirectly on several occasions to a limited extent, only it is hoped that advances in diagnostic technology will soon allow the first direct measurement of neutral density. In addition, $n_i$, which represents the main ion density, is uncertain due to impurities that are always present in tokamaks, and the profile of $E_r$ is not always measured in all tokamaks. Thus, if the unmeasured parameters included in Eq.(5) are different, the value of $Re$ can be different even if the other measured variables are the same, which makes the approach of the H-mode seem arbitrary.

(15) The H-mode is most often followed by an edge-localized mode (ELM), which is a periodically recurring phenomenon: ELMs are a type of magnetohydrodynamic (MHD) instability that arises during the H-mode. The mechanism of ELM itself is a research topic that should be explained in a separate paper, but ELM occurs in H-mode because the laminar flow region in H-mode is vulnerable to MHD instability. In L-mode, this region is homogenized by turbulent flow. The plasma and neutrals are well mixed, and even the inhomogeneous turbulent vortices shown in figure 3 are only a few millimeters in size. However, when the plasma becomes laminar, the neutrals supplied by recycling can be inhomogeneously distributed in the entire magnetic flux, especially if this region corresponds to an effective rational surface where the plasma particles rotate the tokamak many rounds and return to their original position under the influence of magnetic and electric fields, then an MHD instability that breaks down the entire transport barrier at once, which is the ELM. After the collapse, the H-mode is formed again to recover, but since the magnetic structure and $E_r$ change during this period, the conditions for ELM occurrence are formed again after a certain period of time, so this phenomenon recurs periodically. Preventing or controlling ELM is as important to the success of fusion as controlling the H-mode, and can be accomplished by manipulating the electric or magnetic fields in this region. In fact, a technique has been developed to control ELM by utilizing coils that create resonant magnetic perturbations (RMP).

VI. Additional Findings

If the gyro-center shift phenomenon is real, it should be found in other plasmas than tokamaks. The momentum exchange caused by the charge exchange reaction has solved many questions about the origin of the H-mode as well as other fields. Since this report focuses on the origin of the H-mode, we will only briefly discuss these auxiliary findings. Readers who are not interested in phenomena other than the origin of H-mode may skip this chapter.

(1) Reversed rotation of the cathode spot of an arc discharge: In a cylindrical structure with an axial magnetic field, a direct current arc discharge perpendicular to the magnetic field causes the plasma to rotate under the Lorentzian force due to the applied electric and magnetic fields. However, when the discharge pressure and magnetic field are varied, the plasma rotates in the opposite direction to the Lorentz force in certain conditions. This phenomenon was first discovered by Stark in 1903 and remained unexplained for more than 100 years. A leading textbook on arc plasmas published in 1991 stated that "Numerous attempts have been made to solve this problem, but no solution has been found even at a qualitative level" [14]. 104 years after the discovery, it was shown that the separation of charges formed by Eq.(1) is in the same direction as the applied electric field at the anode, but reversed at the cathode, resulting in opposite Lorentzian forces near the cathode [15]. Thus, a new textbook on arc plasma written in 2020 includes an explanation by the gyro-center shift analysis [16]. This is an important example of how problems that have been unsolved for more than 100 years in physics are eventually solved.

(2) Electric field generated in the Earth's ionosphere; Equatorial electrojet (EEJ) and black auroras: The Earth's ionosphere contains plasma generated by charged particles from the Sun. And here also, there is an area where the plasma and below neutrals overlap, and charge exchange reactions occur, but here, rather than pressure or density gradient, the neutrals near the equator move in one direction due to the Coriolis force caused by the heating from the sun and the rotation of the earth, and the gyro-center shift pushes the ions in one direction, creating a vertical electric field, and the electric field creates Hall current (EEJ), and the perturbation of the magnetic field caused by this current made sailors observe the wiggling compass hundreds years ago. At the polar region of the Earth's ionosphere, auroras occur, and in the aurora, a dark region called the black aurora is observed. In the 1990s, a Swedish satellite passed over this region and measured strong electric fields. This electric field was also identified by Eq.(1) as the electric field generated by the plasma pressure and the neutral density gradients [17].

(3) Origin of Bohm diffusion: Although it was explained that the diffusion occurring in tokamaks is a turbulence-driven phenomenon, the first diffusion equation discovered by Bohm was through experiments of a cylindrical arc plasma with a vertical magnetic field [18]. From the plasma particles in the center of the cylinder, ions move to the edges by gyro-center shift, and the cylindrical device is blocked at the top and bottom by electrical conductors, so electrons move through the magnetic field to the conductors and then move toward the edges to join the pre-moved ions. (This short-circuit effect was firstly analyzed by Simon in 1959 [19].) The movement of not only ions but also electrons becomes a plasma flow, and since the main driving force in Eq.(1) is the drift velocity due to the plasma pressure gradient, this flow is expressed as a diffusion, which is proportional to temperature and inversely proportional to the magnetic field. The proportionality factor, however, varies with conditions because it is weakly ionized plasma, and was about 1/13 to 1/40 in Bohm's experimental conditions. In short, the diffusion that Bohm found was caused by a combination of gyro-center shift current and short-circuit effects [20]. In addition to low-temperature plasmas such as arc discharges and semiconductor processing devices, Bohm diffusions are also found in cosmic plasmas such as the supernova remnant.

(4) Intrinsic rotation of tokamak plasma: The phenomenon of spontaneous rotation of the plasma in a tokamak, even in the absence of external momentum input, has been observed in all tokamaks since it was first discovered in 1999, but the reason for this phenomenon was not understood until recently. This phenomenon was found to be caused by a breaking in the balance of electrical force between ions and electrons at the boundary of the tokamak [21]. Tokamaks use toroidal currents for stabilization, so under such a toroidal electric field, electrons and ions are accelerated in opposite directions and the magnitude of their forces is the same, so they should cancel each other out and achieve equilibrium. However, neutrals present in the boundary region receive momentum from the ions through charge exchange reactions and transfer it to the wall of tokamak. The momentum transferred from the electron to the neutral is relatively small; therefore, the force of the ions pushing both the electrons and neutrals simultaneously is weakened comparing that of the electrons pushing only the ions, and the ions eventually rotate in the direction of the electrons.

(5) Violation of quasi-neutrality: Someone who wants to deny all the theories in this paper, may consider quasi-neutrality. Quasi-neutrality is one of the fundamental properties of plasma, which means that even if you create an isolated point charge inside the plasma, the electrons that make up the plasma are free to move around it and nullify it. This shielding of the electrons is called Debye shielding, and it leads to the conclusion that outside of this shielding, the point charge is ineffective, so there can be no electric field inside the plasma. In simple terms, this is similar to the way that an electric field cannot be created inside an electrical conductor, such as a metal, due to the presence of free electrons. However, it turns out that this quasi-neutrality is violated if enough charge exchange reactions occur. [3]. The charge exchange reaction has the effect of creating a point charge because the ions suddenly change their orbits. While there is no electric field outside the Debye shielding, there is an electric field inside it. If enough charge exchange reactions occur, all the space in the plasma becomes the inside of the Debye shieldings, creating an electric field.

VII. The meaning of the physical origin of H-mode is revealed

Commercial fusion energy has been quested for 70 years now, and the phenomenon that has held it back was the diffusion by turbulence, called anomalous transport, and the inability to control it. Eq.(5) makes it clear under what conditions turbulence forms and is suppressed. Based on Eq.(5), if turbulence can be controlled, it means that the density and temperature of the plasma can be easily increased, and the confinement time for such high-pressure plasma can be increased. Since the product of density and confinement time, with high enough temperature, must be above a certain value for the fusion plasma to be able to generate energy on its own (Lawson criterion), the control of turbulence by Eq.(5) is an essential means for the success of nuclear fusion. The plasma analysis using gyro-center shifts will also contribute to the advancement of general plasma physics. In addition to other discoveries mentioned in Chapter VI, it can be used in many plasma applications, including solar flares.

# Appendix 1. Driving Eq.(1) of GCS current

This expression starts from the fact that the Lorentz force is in equilibrium with the force generated by the charge exchange reaction, i.e.

$$\vec{J} \times \vec{B} = force = n_i m_i v_{i-n} \nu_{i-n} \qquad (A\text{-}1)$$

The force on the right-hand side is equal to the change in ion momentum by the charge exchange reaction. Here, the momentum is the product of the ion mass ($m_i$) and the relative velocity between the ion and neutral ($v_{i-n}$), and the force is the product of the momentum multiplied by the number that this reaction occurs per unit time ($\nu_{i-n}$) and the ion density ($n_i$). $v_{i-n}$ is in the direction perpendicular to the magnetic field, as shown in figure 6, and the magnetic field is in the axial direction of the torus. In the above equation, the left-hand side is a vector cross product, so the current direction, the magnetic field direction, and the force direction on the right-hand side must be all perpendicular to each other, so the current J is in the radial direction. Thus, Eq.(1) is derived.

# Appendix 2. Amount of particles diffused in one rotation in a micro-vortex

Specifically, the movement of the particles in the micro-vortex in figure 3 as it rotates from left ([A] in Figure A-1) to right ([B]) and back to left ([C]) in one revolution is shown below.

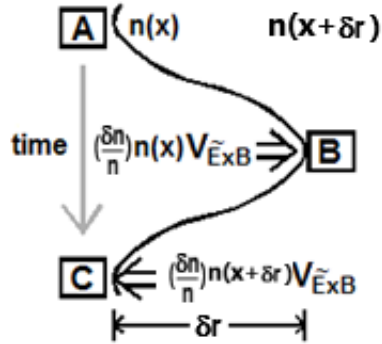

Figure A-1 Diffusion by ExB dirift in the micro-vortex.

In figure A-1, the density n(x+δx) at a distance x to the right by δx can be expressed as n(x)+δrn'. where n' is the gradient of the density (dn/dr). The perturbed component of the density in [A], $(\frac{\delta n}{n})n(x)$, moves toward [B] at the ExB drift speed of the micro-vortex, $v_{\vec{E}\times B}$, and is added to n(x+δx), and then the perturbed component returns to [C]. This process is illustrated in the table below.

|     | x | x + δx |
|-----|---|--------|
| [A] | $n(x) \equiv n$ | $n(x+\delta x) = n + \delta r n'$ |
| [B] | $n - (\frac{\delta n}{n})n$ | $n + \delta r n' + (\frac{\delta n}{n})n$ |
| [C] | $n - \cancel{(\frac{\delta n}{n})n} + \cancel{(\frac{\delta n}{n})n} + (\frac{\delta n}{n})\delta r n' + \cancel{(\frac{\delta n}{n})^2 n}$ $\approx n(x) - (\frac{\delta n}{n})\delta r n'$ | $n + \delta r n' + \cancel{(\frac{\delta n}{n})n} - \cancel{(\frac{\delta n}{n})n} + (\frac{\delta n}{n})\delta r n' - \cancel{(\frac{\delta n}{n})^2 n}$ $\approx n(x+\delta x) + (\frac{\delta n}{n})\delta r n'$ |

n [B], the density at position x+δx multiplied by $\frac{\delta n}{n}$ returns to [C], so if we remove the terms with opposite signs and the term containing $(\frac{\delta n}{n})^2$ in [C] ($\frac{\delta n}{n} \ll 1$), we can see that the change after one rotation is $\frac{\delta n}{n}\delta r \frac{dn}{dr}$. For the case with large density on the left side as shown in figure 3, $n' < 0$. The diffusion of particles described above occurs the same for both left-right-left and right-left-right rotations.

## Acknowledgments

This research was supported by the R&D Program of "KSTAR Experimental Collaboration and Fusion Plasma Research (EN2101-12)" through the Korea Institute of Fusion Energy (KFE) funded by Government.

# References

[1] F. Wagner, G. Becker, K. Behringer, D. Campbell, A. Eberhagen, W. Engelhardt, G. Fussmann, O. Gehre, J. Gernhardt, G. v. Gierke, G. Haas, M. Huang, F. Karger, M. Keilhacker, Q. Kluber, M. Kornherr, K. Lackner, G. Lisitano, G. G. Lister, H. M. Mayer, D. Meisel, E. R. MIiller, H. Murmann, H. Niedermeyer, W. Poschenrieder, H. Rapp, H. Bohr, F. Schneider, G. Sil.ler, E. Speth, A. Stabler, K. H. Steuer, G. Venus, O. Vollmer, and Z. YuF, *Phys. Rev. Lett.* **49**, 1408 (1982)

[2] K. C. Lee, *Phys. Plasmas* **13** 062505 (2006)

[3] K. C. Lee, *J. Korean Phys. Soc.*, **6** 1944 (2013).

[4] TFR Group and A. Truc, *Nucl. Fusion* **26** 1303 (1986)

[5] Ch. P. Ritz, R. V. Bravenec, P. M. Schoch, R. D. Bengtson, J. A. Boedo, J. C. Forster, K. W. Gentle, Y. He, R. L. Hickok, Y. J. Kim, H. Lin, P. E. Phillips, T. L. Rhodes, W. L. Rowan, P. M. Valanju, and A. J. Wootton, *Phys. Rev. Lett*. **62**, 1844 (1989)

[6] K. C. Lee, *Plasma Phys. Control. Fusion* **51**, 065023 (2009).

[7] R. J. Goldston and P. H. Rutherford, *Introduction to Plasma Physics* Bristol IPP 160p (1995)

[8] B. A. Carreras; L. W. Owen; R. Maingi; P. K. Mioduszewski; T. N. Carlstrom; R. J. Groebner, *Phys. Plasmas*, **5**, No. 7, 2623 (1998)

[9] F. Ryter, S.K. Rathgeber, L. Barrera Orte, M. Bernert, G.D. Conway, R. Fischer, T. Happel, B. Kurzan, R.M. McDermott, A. Scarabosio, W. Suttrop, E. Viezzer, M. Willensdorfer, E. Wolfrum and the ASDEX Upgrade team, *Nucl. Fusion* **53** 113003 (2013)

[10] Y. Miura, Y. Asahi, K. Hanada *et al*., *Proceedings of the 16th International Conference on Fusion Energy (IAEA, Montreal, Quebec, 1997)* **1** p.167. (1997)

[11] E. Viezzer, T. Putterich, R. M. McDermott, G. D. Conway, M. Cavedon, M. G. Dunne, R. Dux, E. Wolfrum and the ASDEX Upgrade Team, *Plasma Phys. Control. Fusion* **56** 075018 (2014)

[12] J. Wesson, Tokamaks, 4[th] Edition, New York Oxford UP, 185p (2011)

[13] R. Maingi, R. E. Bell, J. M. Canik, S. P. Gerhardt, S. M. Kaye, B. P. LeBlanc, T. H. Osborne, M. G. Bell, E. D. Fredrickson, K. C. Lee, J. E. Menard, J.-K. Park, S. A. Sabbagh, and NSTX team, *Phys. Rev. Lett*. **105**, 135004 (2010)

[14] Y. P. Raizer, Gas Diacharge Physics, Berlin Heidelberg Springer-Verlag, p262 (1991)

[15] K. C. Lee, *Phys. Rev. Lett.* **99** 065003 (2007).

[16] I. Beilis, Plasma and Spot Phenomena in Electrical Arcs, Springer Nature Switzerland AG, p795 (2020)

[17] K. C. Lee, *Phys. Plasmas* **24** 112505 (2017)

[18] D. Bohm, The Characteristics of Electrical Discharges in Magnetic Fields, A. Guthrie and R. K. Wakerling, Eds. New York, NY, USA: McGraw-Hill, p. 201 (1949)

[19] A. Simon, An Introduction to Thermonuclear Research: A Series of Lectures Given in 1955. New York, NY, USA: Pergamon, (1959)

[20] K. C. Lee, *IEEE Trans. Plasma Sci.* **43** 494 (2015).

[21] K. C. Lee, S. G. Lee, *Current Appl. Phys.* **55** 16 (2023)